\documentstyle[12pt,epsf]{article}
\title{DEMOCRATIC DECAY}
\author{Andrey M. SHIROKOV \thanks{A talk presented
at the conference Perspectives of
Nuclear Physics in the Late Nineties (Hanoi, March 1994).}\\
Institute for Nuclear Physics, Moscow State University, \\
Moscow 119899, Russia \\
E-mail: shirokov@compnet.msu.su}
\date{ }
\begin{document}
\maketitle
\begin{abstract}
Results of some recent investigations of democratic three-body
decays of nuclear systems are discussed. In particular, we consider
experimental studies of three-body ($\alpha+N+N$)-decay of $A=6$ nuclei
and $J$-matrix calculations of monopole excitations in $^{12}$C and $E1$
transitions  in $^{11}$Li in the three-body-continuum cluster models $\alpha
+\alpha +\alpha$ and $^{9}$Li $+\ n+n$, respectively.
\end{abstract}
\renewcommand{\thesection}{\arabic{section}.}
\section{Introduction}
\par
\ \ \ \ Investigation of the decay properties of nuclear
resonant states is an important
part of the nuclear structure studies. Usually in such investigations
only two-body decay channels are allowed for. Three-body
decays are much more informative than the two-body ones, but
the full  description of a three-body decay is a very complicated
problem. Thus it is very important to derive simple but accurate approximate
approaches for the analysis of the experimental data and for making reliable
theoretical predictions.  One of the  approaches of such kind is the
approach based on the democratic decay approximation.
\par
A three-body  continuum spectrum wave function in the asymptotic region
is generally a superposition of the components describing two-
and three-body decay channels~$^1$.  The democratic decay approximation
accounts for the three-body channel only;  two-body channels associated with
the appearance of bound two-body subsystems and 'non-democratic'
subdivision of the system of the type (2+1), are not allowed for in the
approximation. Therefore, this approximation is valid only for the study of a
three-body system in the case when all two-body decay channels are closed
and the only open channel is a three-body channel. Generally speaking, if
the three-body channel is the only open one, than the system can have
two types of asymptotics~$^1$. One of the asymptotics corresponds to the
situation when one of the particles is scattered by another one and the
third particle is a spectator. From the general physical point of view,
this asymptotics is supposed to be of little importance for
a nuclear system excited in some nuclear reaction and decaying via a
three-body channel. The alternative type of the asymptotics
is a superposition of ingoing and outgoing six-dimensional spherical waves
(or, equivalently, six-dimensional plane wave and ingoing or outgoing
six-dimensional spherical wave) and
corresponds to the situation when the decaying system emits (or/and absorbs)
three particles from one and the same point in space. The version of the
scattering theory that allows only for three-body asymptotics of this type
is called true three-body scattering theory, or, alternatively, $3 \to 3$
scattering theory; the approximation based on the allowance for  $3
\to 3$ scattering only in the examination of nuclear reactions is called
democratic decay approximation.
\par
Nuclear structure studies within the framework of democratic decay
approximation have been started in 70th by Prof.\ R.I.Jibuti with
collaborators.  The reviews of the results obtained by the Tbilisi group can
be found in refs.\ $^{2,3}$. Recently a considerable progress
has been made in both
theoretical and experimental studies of democratic decays. Below we shall
briefly review some of the recent results. In particular, we shall discuss
the results of experimental studies of democratic decays of $A=6$ nuclei via
the three-body channel $\alpha +N+N$, theoretical investigations of monopole
excitations of $^{12}$C nucleus in the cluster model
$\alpha +\alpha +\alpha$ and  $^{11}$Li structure  studies in the
cluster model $^9$Li$+n+n$.
\section{Democratic decay of $A=6$ nuclei }
\ \ \ \ Detailed experimental study of the three-body decay of $T=1$,
$J^{\pi}=0^+$ and $2^+$
states in $^6$He, $^6$Li and $^6$Be nuclei has been performed by the
group from Kurchatov Institute (see~$^{4,5}$ and references therein).
The democratic decay concept has been for the first time experimentally
tested in this investigation, and the
democratic nature of the decay of these states via the only open channel
$\alpha + N + N$ has been unambiguously demonstrated.
\par
The wave function in the democratic  decay approximation has the following
asymptotics:
\begin{equation}
\Psi^- \sim \kappa^{-1/2}\left[ (2\pi)^{-3}
\exp (i\mbox{\boldmath $r_{ab}q_{ab}$}+i\mbox{\boldmath $r_{ab,c}q_{ab,c}$})
 + F \frac{e^{-i\rho\kappa}}{\rho ^{5/2}}\right]\chi_{SM_S}\ ,
                         \label{1}
\end{equation}
where the first term is a six-dimensional plane wave while the second one is
an ingoing six-dimensional spherical wave; $F$ is the $3 \to 3$ scattering
amplitude; $\mbox{\boldmath $r$}_{ab}$ and $\mbox{\boldmath $r$}_{ab,c}$
are Jacobi coordinates, and $\mbox{\boldmath $q$}_{ab}$ and
$\mbox{\boldmath $q$}_{ab,c}$  are associated canonically conjugated
momenta; hyper-radius $\rho =
(\mbox{\boldmath $r$}_{ab}^2 + \mbox{\boldmath $r$}_{ab,c}^2)^{1/2}$,
and $\kappa =
(\mbox{\boldmath $q$}_{ab}^2 + \mbox{\boldmath $q$}_{ab,c}^2)^{1/2}
=(2mE)^{1/2}/\hbar$; $E$ is the energy measured from the three-body breakup
threshold; $S$ and $M_S$ are the total spin of the system and its
projection while $\chi_{SM_{S}}$ is the spin part of the wave function.  It
is natural to expand the decay amplitude $F$ in hyperspherical
harmonics $Y_{\Gamma}\equiv Y_{KLM_L}^{\gamma}$ and to couple the
three-body orbital momentum $L$ and the spin $S$ into the total angular
momentum $J$.  As a result, for the decay amplitude $F_{JM}$ of a resonant
state with the total spin $J$ and its projection $M$ we have the following
expression~$^{3,6}$:
\begin{equation}
F_{JM} = \sum_{\Gamma} f_{KLS}^{\gamma}\cdot\left[
Y_{KLM_L}^{\gamma}\chi_{SM_{S}} \right]_{JM}\ .                    \label{2}
\end{equation}
In eq. (\ref{2}), $K$ is hypermomentum, the multi-index $\Gamma \equiv
\{K,L,M_{L},S,\gamma\}$ plays the role of a channel index for the $3\to 3$
scattering, and the multi-index $\gamma$ stands for the set of all
additional quantum numbers labelling different channels with the same
quantum numbers $K,L,M_{L},S$. It is convenient to use the Yamanouchi symbol
$r$ and  the Young scheme [$f$] for $\gamma$ if there are three identical
particles in the final state; otherwise orbital momenta $l_{ab}$ and
$l_{ab,c}$ corresponding to Jacobi coordinates $\mbox{\boldmath $r$}_{ab}$
and $\mbox{\boldmath $r$}_{ab,c}$ are usually used for $\gamma$.
\par
Eq. (\ref{2}) is a  generalization  of the usual angular momentum
decomposition of a two-body scattering amplitude and can be formally
applied to any three-body decay. Note, that hypermomentum $K$
is not an integral of motion, and due to the two-body interactions
components labelled by different values of
$K$ mix together in  the wave function, so,
partial amplitudes $f_{\Gamma} \equiv f_{KLS}^{\gamma}$
with various values of $K$ are present in  the
expansion (\ref{2}). For a  two-body decay channel the expansion
(\ref{2}) is non-convergent;  in the case  when two-body subsystems of
the decaying three-body system have sharp resonances, the convergence can
appear to be very slow. If two-body subsystems of the three-body system
have neither bound  nor sharp resonant states, than the $3\to 3$ scattering
amplitude $F_{JM}$  converges rapidly and only few terms play an important
role in eq.  (\ref{2}). Thus only in the case of democratic decay of such
system it is possible to extract  partial
amplitudes $f_{\Gamma} \equiv f_{KLS}^{\gamma}$ from
the experimental data, i.e., to solve the problem
that is somewhat similar to the problem of the spin assignment to a
resonant state by the analysis of two-body scattering data.
\par
 The hyperspherical harmonics $ Y_{KLM_L}^{\gamma}$ are  functions of
five angles in the formal six-dimensional momentum space
\{$\mbox{\boldmath $q$}_{ab},\mbox{\boldmath $q$}_{ab,c}$\}. One of the
angles, $\vartheta = \arctan (q_{ab}/q_{ab,c})$, rules the energy
distribution between two Jacobi coordinates. Therefore, in order to
evaluate partial amplitudes $f_{KLS}^{\gamma}$, one should analyze energy
spectra of emitted particles. The analysis of ref.~$^4$ of the three-body
decay $^6$Be$(0^+)\to \alpha+p+p$ will serve us as an example.
\par
$^6$Be levels have been populated in the reaction
$^6$Li($^3$He,$^3$H)$^6$Be at $E_{^{3}He}=40$~MeV. $^3$H has been
detected in the coincidence with $\alpha$
and/or $p$ from the decay $^6$Be $\to \alpha +p+p$. The detection of $^3$H
of appropriate energy made it possible to select the levels of interest.
Taken from ref.~$^4$ the
spectrum of $\alpha$-particles from the decay of the
$^6$Be $0^+$ state  in the beryllium rest frame is depicted on fig.~1.
\begin{figure}[tb]
\epsfverbosetrue
\epsfysize=9cm \epsfbox{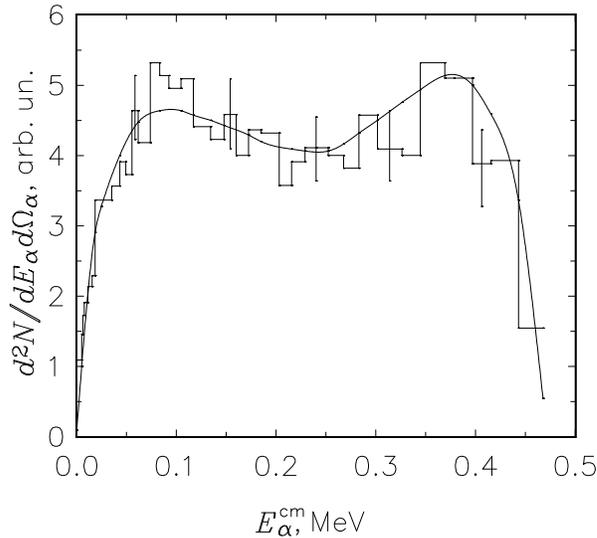}
\caption{The $\alpha$-particle spectrum (histogram) from $^6$Be($0^+$)
decay in the beryllium rest frame. The solid line is the hyperspherical
harmonics description of the data. The figure is taken from ref.~$^4$.}
\end{figure}
The peculiar form of the spectrum cannot be reproduced in the model of
independent escape of the decay products (phase volume calculations), nor
can it be reproduced in the model of sequential decay via the resonant
states of  $p$-$p$ or $\alpha$-$p$ subsystems. At the same time, as
is seen from fig.~1, the data are
perfectly reproduced using expansion (\ref{2}) and
allowing only for the terms with $K=0$ and 2. Nevertheless, the set of the
partial amplitudes \{$f_{KLS}^{\gamma}$\} cannot be uniquely
determined from the analysis of
the $\alpha$-spectrum because various sets of
\{$f_{KLS}^{\gamma}$\} give the fit to the data of the same quality. One
should involve in the analysis the energy spectrum of protons from the same
reaction. The predictions for the proton spectrum obtained using various
sets of  \{$f_{KLS}^{\gamma}$\} fitted to the $\alpha$-spectrum, differ
significantly. From the fit to both $\alpha$- and $p$-spectra,
the relative weights of
($6\pm 5$)\%; ($44\pm 12$)\%; and ($50\pm 17$)\% for the components $K=0$;
$K=2,\ S=0$; and $K=2,\ S=1$, respectively, have been obtained in ref.~$^4$.
\par
The set of partial amplitudes  \{$f_{KLS}^{\gamma}$\} can be used for the
predictions of the kinematically complete experiment based on the
triple-coincidence $^3$H $+p+\alpha$ measurements. As is seen from fig.~2,
\begin{figure}[tb]
\epsfverbosetrue
\epsfysize=9cm \epsfbox{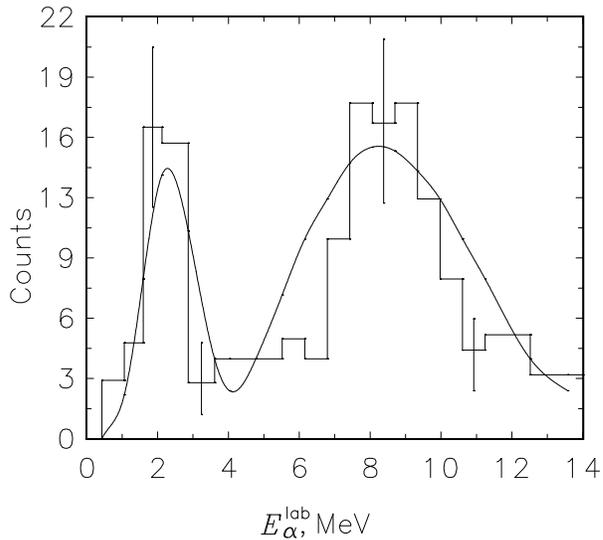}
\caption{The histogram is the triple coincidence  distribution
measured in the kinematically complete experiment; the solid line is the
prediction based on the $\alpha$ and proton spectra calculated using
hyperspherical harmonics expansion. The figure is taken from ref.~$^4$.}
\end{figure}
the set of \{$f_{KLS}^{\gamma}$\} fitted to $\alpha$- and p-spectra,
perfectly matches the triple-coincidence data. Note, that the kinematical
conditions of the triple-coincidence
experiment~$^4$ have been chosen in such a way, that the
contribution of the $S=1$ decay has been strongly suppressed. Therefore, the
fine agreement between the theoretical curve and the experimental data on
fig.~2 confirms the good quality of the determination of the relative
weights of $K=0$ and $K=2,\ S=0$ components only.
\section{Democratic decay of $0^+$ states in $^{12}$C}
\ \ \ \ In the previous section we have discussed the application of the
democratic decay approximation to the analysis of experimental data. Now we
shall discuss the use of the true three-body scattering theory in
calculations.
\par
It is well known that the giant monopole resonance generates a
well-pronounced peak of the monopole
excitation strength in heavy nuclei exhausting
an essential fraction of the energy-weighted sum rule (EWSR). In light
nuclei, the giant monopole resonance is very weak
and exhausts, e.g., in $^{12}$C, only $\sim 5$\% of EWSR~$^7$.
However, the strong peak of $E0$ strength exhausting $\sim 90$\% of
EWSR has been predicted in various microscopic calculations (see, e.g.,
ref.~$^8$). It has been shown in ref.~$^9$ that this discrepancy should be
attributed to the continuum spectrum effects, i.e., to the decay properties
of the giant monopole resonance.
\par
The 3$\alpha$ cluster model has been used in ref.~$^9$. The ground state
wave function has been obtained by diagonalization of the three-body
Hamiltonian using the harmonic oscillator basis.
For $0^+$ continuum spectrum states only the decay channel
$^{12}$C $\to \alpha + \alpha +\alpha$ has been allowed for. This channel
is the  single open channel for  low-lying $0^+$ states and is
supposed to dominate for  $0^+$ states with excitation energies of up to
$\sim 20$~MeV. Continuum spectrum wave functions have been
calculated by means of the oscillator representation of scattering theory
($J$-matrix method)~$^{10}$ which had been generalized on the case of
the true few-body scatting in ref.~$^{11}$.
\renewcommand{\thefootnote}{\fnsymbol{footnote}}
\setcounter{footnote}{1}
\par
In the $J$-matrix approach, the continuum spectrum wave function
$\psi^{\gamma}_{K}$ for a given
three-body decay channel $\Gamma \equiv \{K,\gamma \}$\footnote{Hereafter
we shall omit the spin part of the wave functions and
 quantum numbers $S$, $L$, and $M_L$ in the channel index
$\Gamma \equiv \{K,L,M_{L},S,\gamma\}$ and in the indexes of the functions
used. In the case of spinless particles (e.g., $\alpha$-particles), $S=0$,
$L$, and $M_L$ are integrals of motion, therefore, all functions are
labelled with the same values of $S$, $L$, and $M_L$ and channels
$\Gamma \equiv \{K,L,M_{L},S,\gamma\}$  and
$\Gamma'\equiv \{K',L',M'_{L'},S',\gamma'\}$ with $L \neq L'$ and/or $M
\neq M'_{L'}$ are not coupled.}  is
expressed as an infinite series expansion in six-dimensional harmonic
oscillator functions $\varphi_{n\,K}$,
\begin{equation}
\psi^{\gamma}_{K} = \sum_{n=0}^{\infty} \;
D^{\gamma}_{n\,K}(E) \ \varphi_{n\,K}\,.
\label{3}
\end{equation}
The basic approximation of the $J$-matrix approach is in neglecting
potential energy matrix elements $V_{N\Gamma}^{N'\Gamma'} \equiv
\left( V_{12}+V_{13}+V_{23} \right)_{N\Gamma}^{N'\Gamma'}$ with a large
number  of oscillator quanta $N=2n+K>\tilde{N}$ and/or $N'=2n'+K'>\tilde{N}$;
the infinite kinetic
energy matrix $T$ is not truncated. The exact wave functions of the
Hamiltonian with the infinite kinetic energy matrix and truncated potential
energy matrix for any positive energy $E$ can be found  by algebraic
methods.  To do it, one should first diagonalize the truncated Hamiltonian
matrix, i.e., calculate its eigenvalues $E_{\lambda}$ and corresponding
eigenvectors $\zeta_{n\Gamma}^{\lambda}$. Next the matrix elements
${\cal P}^{n',\,\Gamma'}_{n,\,\Gamma}(E)$ of the matrix ${\cal P}$,
\begin{equation}
{\cal P}^{n',\,\Gamma'}_{n,\,\Gamma}(E)
= \sum_{\lambda}\frac {\zeta^{\lambda}_{n'\,\Gamma'} \;
\zeta^{\lambda}_{n\,\Gamma} } { E_{\lambda}\;-\;E } \; ,
\label{4}
\end{equation}
are calculated. The $3\to 3$ scattering $S$-matrix can be easily computed
by  the expression
\begin{equation}
S = (A^{(+)})^{-1} A^{(-)}\; ,   \label{5}
\end{equation}
where matrix elements of matrices $A^{(+)}$ and $A^{(-)}$,
\begin{equation}
A^{(\pm)}_{\Gamma\,\Gamma'} =
\delta_{\Gamma\,\Gamma'}\;C^{(\pm)}_{\nu,\,K}(E)
\ + \ {\cal P}^{\nu',\,\Gamma'}_{\nu,\,\Gamma}(E)\;
T^{\Gamma'}_{\nu',\,\nu'+1}\;C^{(\pm)}_{\nu'+1,\,K'}(E)\;;
\label{6}
\end{equation}
$\nu=\frac{1}{2} (\tilde{N}-K)$ and
$\nu'=\frac{1}{2}(\tilde{N}-K')$ are  the truncation boundaries of the
Hamiltonian matrix in the channels $\Gamma\equiv \{K, \gamma \}$ and
$\Gamma'\equiv \{K', \gamma'\}$, respectively;
the kinetic energy matrix elements
$T^{\Gamma}_{n,\,n+1} =
 -\frac{1}{2}\hbar\omega \; \sqrt{(n+1)(n+K+3)}$;
the eigenvectors $C^{(+)}_{n,\,K}(E)$ and $C^{(-)}_{n,\,K}(E)$
of the infinite kinetic energy matrix T
can be easily calculated by
analytical expressions or by recurrent relations (see ref.~$^{11}$
for details).
\par
Finally, for the wave function with outgoing six-dimensional spherical wave
in the channel ${\Gamma}'$ and ingoing six-dimensional spherical
waves in all allowed channels $\Gamma$, the expansion
coefficients $D^{(\Gamma')}_{n\, \Gamma}(E) \equiv D^{\gamma}_{n\,K}(E)$
[see eq.  (\ref{3})] in the asymptotic region $n \ge \nu$ are
calculated by the expression
\begin{equation}
D^{(\Gamma')}_{n\, \Gamma}(E)
= \frac{1}{2} [ \; \delta_{\Gamma\,\Gamma'} \;
C^{(+)}_{n\,K}(E) \ -\ S_{\Gamma\,\Gamma'}\;
C^{(-)}_{n\,K}(E) \;] \ ,                               \label{7}
\end{equation}
while for calculation of $D^{(\Gamma')}_{n\, \Gamma}(E)$ in the interaction
region $n \le \nu$ the expression
\begin{equation}
D^{(\Gamma')}_{n\, \Gamma}(E) = -
\sum_{\Gamma^{\prime\prime}}\;
{\cal P}^{n\,\Gamma}_{\nu^{\prime\prime}\,\Gamma^{\prime\prime}}(E)\;
T^{\Gamma^{\prime\prime}}_{\nu^{\prime\prime},\,
\nu^{\prime\prime}+1}\;
D^{(\Gamma')}_{\nu^{\prime\prime}+1,\, \Gamma^{\prime\prime}}(E)
\label{8}
\end{equation}
is used.
\par
Ali-Bodmer $\alpha\alpha$ potential $^{12}$ has been used in ref.~$^9$;
Coulomb interaction has been neglected. The truncation boundary
$\tilde{N}=18$ has been used in the calculations, i.e., the truncated
Hamiltonian matrix accounted for all $3\alpha$ oscillator states up to
$18\hbar\omega$. The convergence for both the ground and continuum $0^+$
states is achieved at $\tilde{N} \approx 18$ provided that $\hbar \omega
=10$ MeV. In the asymptotic
region $n \ge \nu$ all channels $\Gamma$ with  $K \leq 10$
have been allowed for. However, in the interaction region
$n \le \nu$ we have accounted
for all oscillator states up to $18\hbar\omega$, thus the hyperspherical
harmonics with $K \leq 18$ have been allowed for in the interaction region.
The inclusion of components with $K \geq 10$ in the channel space have
not produced changes in the results.
\par
The results of  the calculations of ref.~$^9$ are presented on fig.~3.
\begin{figure}[tb]
\epsfverbosetrue
\epsfxsize=13cm \epsfbox{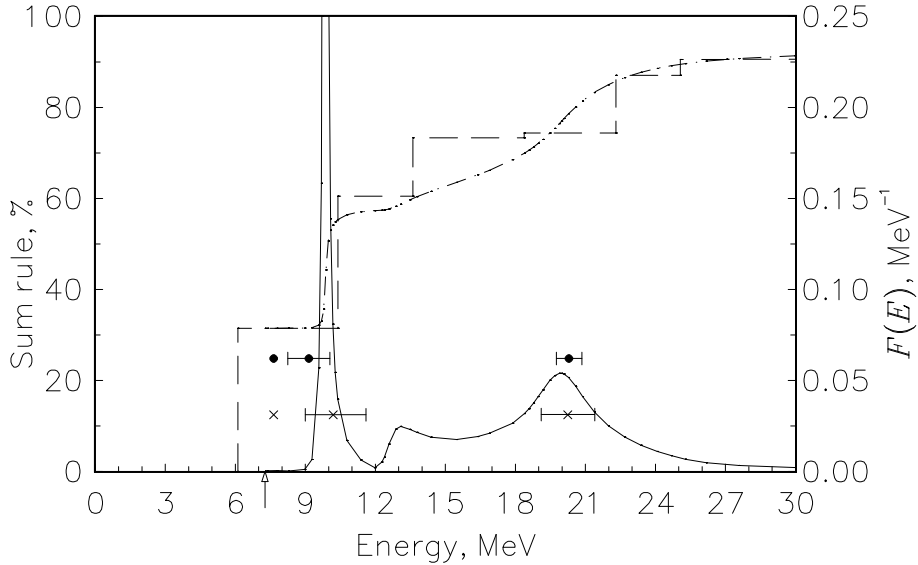}
\caption{The solid line displays the monopole strength function
$F(E)=E\left| M_0 \right|^2/\sigma_{\infty}$ for the states in continuum.
The dash-dotted line is a plot of the  function $S(E)$ [see eq.\
(\protect\ref{9})], i.e., it displays the fraction of monopole EWSR
exhausted by the  states  with excitation energies less than $E$. The
dashed line is the same, but obtained by neglecting the effect of
continuum. The experimental energies and the widths of the $E0$ resonances
are indicated by crosses~$^7$ and circles~$^{13}$. The arrow indicates
the $^{12}$C $\to 3\alpha$ reaction threshold. The figure is taken from
ref.~$^9$.}
\end{figure}
The solid line in fig.~3 represents the $E$-dependence of the
$E0$ transition strength $F(E)=E \left| M_0 \right| ^2 /\sigma_{\infty}$
to the states above the $^{12}$C $\to 3\alpha$  threshold. Here, $ M_0$ is
the matrix element of the $E0$ transition from the ground state to the
states with energy $E$. The dash-dotted curve displays the sum of the
contributions to EWSR from all states with excitation energies less than
$E$, i.e., the function
\begin{equation}
S(E)=\frac{\sigma(E)}{\sigma_{\infty}} \times 100\% \ , \label{9}
\end{equation}
where
\begin{equation}
\sigma(E)= \int_{0}^{E} E' \left| M_0 \right| ^2 dE'\ ,
\label{10}
\end{equation}
the monopole sum rule
$\sigma_{\infty} \equiv \sigma(\infty) = \frac{2\hbar^2}{m}
\left\langle R^2 \right\rangle_0$, and
$\left(\left\langle R^2 \right\rangle_0\right)^{1/2}$ is the rms radius of
the $^{12}$C nucleus. The dashed step-like line is the plot of the function
$S(E)$ calculated disregarding  the effect of the $3\alpha$ continuum. In
this case, the height of each step is equal to the
contribution to the EWSR from the corresponding
quasistationary state obtained by the
diagonalization of the truncated Hamiltonian matrix. It is seen that the
most essential contributions to EWSR  are connected with the
quasistationary levels
with excitation energies 6.11, 10.39, 13.60 and 22.30~MeV
(31\%, 29\%, 13\% and 13\%, respectively). All the remaining levels
with excitation energies up to 35 MeV exhaust about 7\% of the EWSR.
\par
The $^{12}$C binding energy in calculations of ref.~$^9$ has been found to
be $E_B =7.27$~MeV (the experimental value is 7.27~MeV).
The first excited $0^+$ state with the excitation energy $E_1 = 6.11$~MeV
belongs to the discrete spectrum, so, the continuum does not affect its
properties. The experimental value for $E_1$ is 7.66~MeV and this state
belongs to the $3\alpha$ continuum. However, its width is very small
($\Gamma= 9$~eV) and its properties are rather close to those of the
discrete spectrum states.
\par
The allowance for the continuum slightly changes the properties of the
10.39~MeV
$0^+$ state. Namely, the $F(E)$ maximum shifts down to $E=9.9$ MeV,
the state gets some  width and its contribution to EWSR decreases to 26\%.
\par
However, in the 13--25 MeV excitation region the effect of the continuum
appears to be essential. The levels in this excitation energy region get
considerable widths and merge each other, forming very broad (of
$\sim 12$~MeV in width) plateau of a peculiar shape, i.e., the strength
function $F(E)$ is almost constant between 13 and 25~MeV except for a
bump of the width of 2--3~MeV at an  excitation energy of about 20~MeV.
The bump should be related to the 20~MeV monopole resonance observed in
refs.~$^{13,7}$. 33\% of the EWSR is exhausted in the 12--26~MeV excitation
energy region, with 16\% of the EWSR being exhausted in the 18--22~MeV
interval.
\par
The approach, as is seen from fig.~3, reproduces quite correctly the
positions of the strong $E0$ absorption regions. The broad and smooth
pedestal of the 20~MeV $E0$ excitation strength maximum can be easily
misinterpreted in experimental studies as a background. The shape of the
background that has been subtracted by Eyrich et al~$^7$ in their study of
the 20~MeV resonance, is similar to the shape of the pedestal on fig.~3. If
the experimental background is a real manifestation of the smooth pedestal
of the $E0$ strength, than it should be higher for small
scattering angles. From fig.~3 of ref.~$^7$ it is seen,
that the background is really
enhanced at small angles. So, the experimental spectra of ref.~$^7$ is in
qualitative agreement with theoretical results of ref.$^9$. The further
investigation of the nature of the background is needed.
\par
The results presented on fig.~3 give the explanation why the giant monopole
resonance is nearly not seen in $^{12}$C. There are two physical
reasons. First one is the strong fragmentation of the $E0$ strength in
$^{12}$C. As a result, more than 50\% of the EWSR is exhausted by the
7.65~MeV and 10~Mev resonances, and not too much of the $E0$  strength is
left for the giant monopole resonance. The second reason is the continuum
spectrum effect. Due to the democratic decay, the giant monopole resonance
gets a large width and merge with other $0^+$states. As a result, the
$E0$ strength becomes very smooth, and the giant monopole resonance
manifests itself only as a non-prominent bump on the smooth pedestal.
About 15\% of the EWSR is exhausted in the vicinity of 20~MeV bump, but if
the pedestal is treated as a background, than the contribution of the
resonance to the EWSR is severely underestimated.
\section{$^{11}$Li structure and the soft dipole mode}
\par
\ \ \ \ Light neutron-rich nuclei like $^6$He, $^{8}$He, $^{11}$Li,
$^{14}$Be, etc., are extensively studied now (see  recent
review papers~$^{14,15}$ and   references therein). These nuclei are often
called exotic nuclei; they are also often called Barromean nuclei, because
they have a well-pronounced cluster structure $core + n + n$ and
after removing a neutron from any of these nuclei, the remaining system
$core + n$  becomes unbound.\footnote{The heraldic symbol of the dukes of
Barromeo is three rings connected in such a manner that after  removing
of any of the rings the two others become disconnected.} Two excess
loosely-bound neutrons have a wide spatial distribution with long tail
reaching far out and form the so-called neutron halo.   The neutron halo
manifests itself in  a number of unusual properties of exotic nuclei, in
particular, in a large value of the rms radius of the nucleus, in the
existence of the soft dipole mode, etc.
\par
It is natural to make use of the democratic decay approximation in the
study of Barromean nuclei. In ref.~$^{16}$, $^{11}$Li has been studied in
the cluster model $^9$Li $+\ n+n$. It has been shown in ref.~$^{16}$ that
the democratic decay approximation within the framework of
the $J$-matrix approach described in the previous section, is advantageous
for the investigations not only of  the continuum states, but of the
ground state as well.
\par
In a usual variational calculation with the oscillator basis, it is
impossible  to describe the slowly-dying asymptotic
tail of the halo neutron wave functions. In order to
describe the neutron halo properties of the $^{11}$Li nucleus, one
is pushed to account for the oscillator states with a very large
number of oscillator quanta. In the $J$-matrix approach, the infinite
number of the oscillator states can be allowed for. For a bound state, the
expansion coefficients $D_{n\, \Gamma}(E)
\equiv D^{\gamma}_{n\,K}(E)$ [see eq.  (\ref{3})] in the asymptotic region
$n \ge \nu$ are of the form$^{11}$:
\begin{equation}
 D_{n\, \Gamma}(E) = \alpha_{\Gamma}\; C^{(+)}_{n\,K}(E) \,.
                    \label{11}
\end{equation}
The bound state asymptotic normalization constants, $\alpha_{\Gamma}$,
are calculated by handling  numerically  the set of linear
homogeneous equations$^{11}$
\begin{equation}
\sum_{\Gamma'}\,A^{(+)}_{\Gamma
\Gamma'}\;\alpha_{\Gamma'} = 0 \; .  \label{12}
\end{equation}
Eq.~(\ref{12}) can be solved if only
\begin{equation}
\det A^{(+)} = 0\,.     \label{13}
\end{equation}
Using eq.~(\ref{13}) the energy of a bound state can be easily calculated.
This method of calculation of the bound state energies is equivalent to the
location of the $S$-matrix pole$^{11}$ corresponding to the bound state, as
is easily seen from eq.~(\ref{5}).
\par
The expansion coefficients
$D_{n\, \Gamma}(E) \equiv D^{\gamma}_{n\,K}(E)$ for  the inner region
$n \leq \nu$  are obtained using eq.~(\ref{8}). Finally, the wave
function should be normalized.
\par
The calculations$^{16}$ have been  performed using the shallow Gaussian
$n$--$^9$Li and $n$--$n$ potentials of ref.~$^{17}$.
Only the states with the spin of the halo-neutron pair $S=0$ have been
considered. The three-body angular momentum of  the ground state of
$^{11}$Li have been supposed to be equal to $L=0$.
The parameter $\hbar\omega = 7.1$~MeV has been used in the calculations.
This value corresponds approximately to the minimum of
the ground state energy $E_{0}$.
\par
The results for the $^{11}$Li ground state for various values
of the truncation parameter $\tilde{N}$ are presented in the table 1.
The variational
ground state energies,  $E^{(d)}_{0}$, obtained by the pure
diagonalization of the truncated Hamiltonian matrix are listed
in the second column, while the $J$-matrix results, $E_{0}$, which are the
solutions of the eq.~(\ref{13}) are listed in the third column.
\begin{table}[tbh]
\centering
\caption{$^{11}$Li ground state properties (see text for details).}
\begin{tabular}{|c|c|c|c|c|} \hline \hline
                     &     \multicolumn{2}{c|}{Ground state energy,}      &
\multicolumn{2}{c|}{Neutron halo}                      \\
 Truncation          &     \multicolumn{2}{c|}{MeV}                       &
\multicolumn{2}{c|}{mean square radius}                \\
 boundary $\tilde{N}$&     \multicolumn{2}{c|}{}                          &
\multicolumn{2}{c|}{$<r^{2}>^{1/2}_{11}$, fm}          \\ \cline{2-5}
                     &       $E_{0}^{(d)}$        &        $E_{0}$        &
$<r^{2}>^{1/2 \ (d)}_{11}$   &   $<r^{2}>^{1/2}_{11}$   \\ \hline
   12                &          -0.012            &         -0.150        &
             2.83            &         3.31             \\
   16                &          -0.116            &         -0.199        &
             2.91            &         3.29             \\
   20                &          -0.171            &         -0.225        &
             2.98            &         3.31             \\
   24                &          -0.202            &         -0.240        &
             3.04            &         3.32             \\ \hline
 Experiment          &   \multicolumn{2}{c|}{-0.247$\pm$0.080}            &
  \multicolumn{2}{c|}{3.16$\pm$0.11}                    \\ \hline \hline
\end{tabular}
\\
\caption{The effect of the Lanczos smoothing on the ground state energy of
$^{11}$Li (see text for details).}
\begin{tabular}{|c|c|c|c|c|c|c|c|c|} \hline \hline
$\tilde{N}$     & 10    & 12    & 14    & 16    & 18    & 20    & 22    & 24
\\ \hline
$E_0$           &-0.119 &-0.150 &-0.181 &-0.199 &-0.215 &-0.225 &
&-0.240 \\ \hline
$E_{0}^{(w.s.)}$&-0.136 &-0.266 &-0.204 &-0.262 &-0.232 &-0.262 &-0.249
&-0.263 \\ \hline \hline
\end{tabular}
\\
\end{table}
The values of the $^{11}$Li r.m.s.\ radius,
$<r^{2}>_{11}^{1/2 \ (d)}$ and
$<r^{2}>_{11}^{1/2}$, obtained by the
pure diagonalization of the truncated Hamiltonian matrix and
with the allowance for the asymptotic tail of the wave function,
respectively, are presented in the 4-th and the 5-th columns of the table.
It is seen, that by locating the $S$-matrix pole using eq.~(\ref{13}) that
is equivalent to the allowance for the long asymptotic tail of the wave
function, the convergence of the ground state is essentially improved.
\par
The results presented in the table 1 have been obtained using Lanczos
smoothing of the
three-body potential energy matrix$^{18}$. The
smoothing improves the convergency of calculations of the
scattering characteristics. Nevertheless, the smoothing causes the
underestimation of the binding energy of a three-body system. The
$^{11}$Li ground state energy obtained in the $J$-matrix calculations
without the use of the smoothing,  $E_{0}^{(w.s.)}$,
shows a staggering as
$\tilde{N}$ increases\footnote{Note, that $J$-matrix calculation is
not a variational one, so,
the ground state energy may behave non-monotonically as
$\tilde{N}$ increases.} (see table~2). $E_{0}^{(w.s.)}$ is seen to converge
to the value of $-0.263$~MeV. The
smoothing results in a usual-type monotonic decrease of $E_0$ as
$\tilde{N}$ increases, but $E_{0}^{(w.s.)} < E_{0}$ for all values
of $\tilde{N}$. At the same time,
the ground state wave functions obtained with the smoothing and without
it, differ very
slightly. So, below we shall discuss only the results of calculations
with the smoothed potential energy matrix.
\par
The quality of the convergence of the wave function in the $J$-matrix
approach can be demonstrated by the plots of momentum distributions.
Figure 4 presents the transverse momentum distribution of
the cluster $^{9}$Li in $^{11}$Li that is
currently supposed (the so-called Serber reaction mechanism)  to be
proportional to the $^{9}$Li transverse momentum distribution
$\frac{dN}{dp_{\bot}}$ in the fragmentation of high-energy $^{11}$Li beams
on target nuclei. The experimental data for $^{11}$Li fragmentation$^{19}$
are also presented on the figure. The results of $J$-matrix
calculations with $\tilde{N}=16$ and $\tilde{N}=24$ are so close to each
other that their plots on fig.4 coalesce and cannot be distinguished.
\begin{figure}[tb]
\epsfverbosetrue
\epsfbox{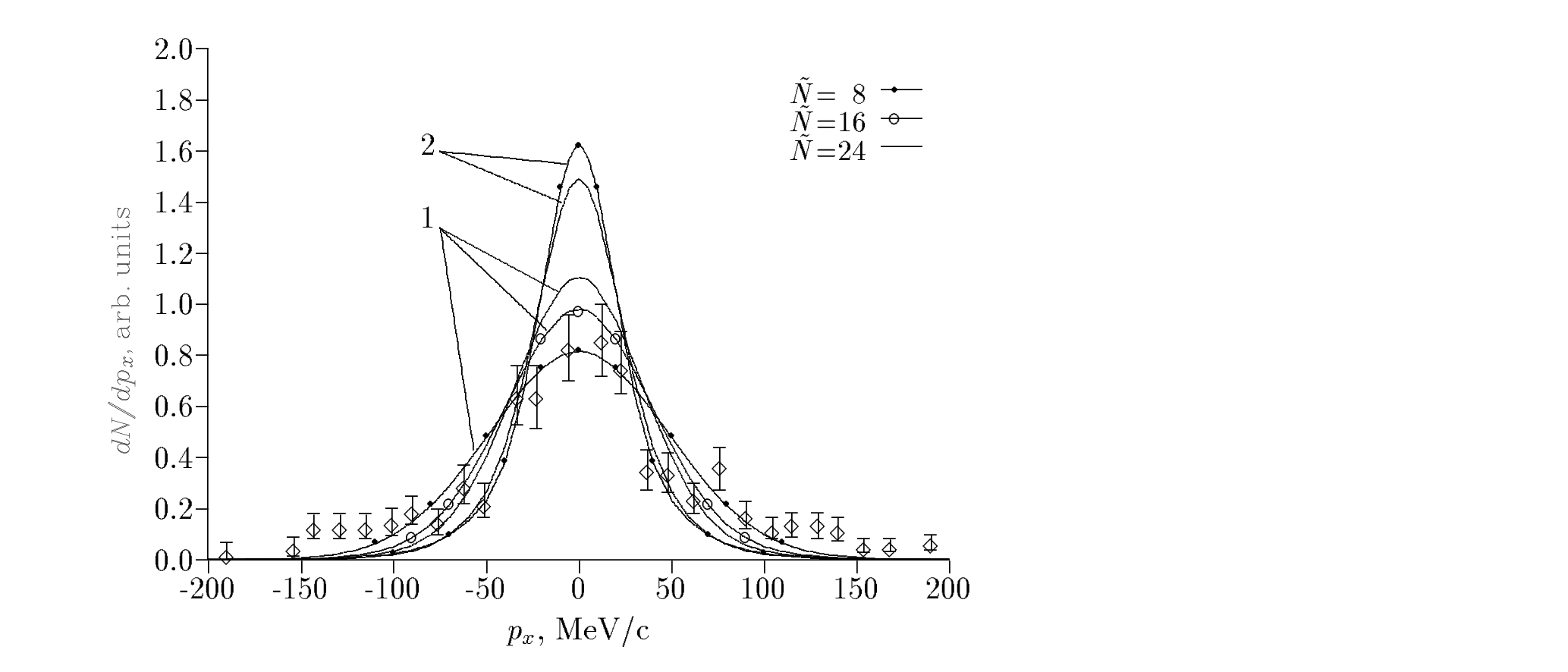}
\caption{The $^{9}$Li transverse momentum distribution in the ground state
of $^{11}$Li. 1 --- variational calculations, 2 --- $J$-matrix calculations.
Experimental data for the $^{9}$Li transverse momentum
distribution in the fragmentation of $^{11}$Li 800 MeV/nucleon beam are
taken from ref.~$^{19}$.}
\end{figure}
\par
It is seen that in
calculation of the ground state, the allowance for the wave function
asymptotics is very important for a weekly-bound system like $^{11}$Li. The
terms of expansion (\ref{3}) with the number of  oscillator quanta
$N\simeq 100$ that cannot be obtained in the usual oscillator-basis
variational calculations, play an essential role in the formation of the
transverse momentum distribution, r.m.s.\ radius, etc.  The convergence of
$<r^{2}>^{1/2}_{11}$, transverse momentum distribution and other properties
of the wave function (e.g., of the weights of its components)  in the full
$J$-matrix calculation is very good. Nevertheless, it is seen that the
r.m.s.\ radius converges to a value that is somewhat larger than the
experimental one, and the calculated transverse momentum distribution is
narrower than the experimental one. These shortcomings can be  overcome by
the adjustment of $n$--$^9$Li potential. We have not aimed to fit the
potential to the $^{11}$Li properties, we have just
take its parameters from ref.~$^{17}$. The differences
between the calculated and
experimental values of r.m.s.\ radius arise from the fact that in
ref.~$^{17}$ the potential has been fitted to the $^{11}$Li
r.m.s.\ radius and the binding energy in the
{\em variational calculation} that
is unable to reproduce these quantities with high accuracy.
\begin{figure}[tbh]
\epsfverbosetrue
\epsfbox{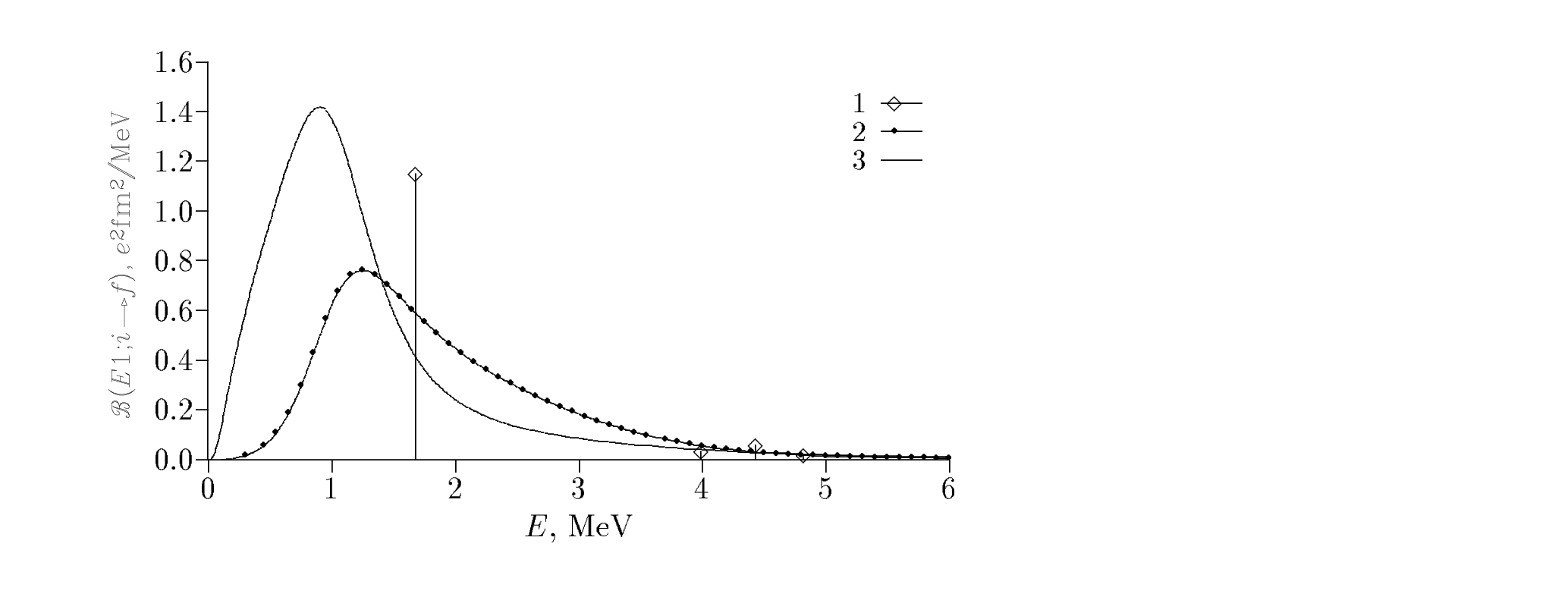}
\caption{Cluster ${\cal B}(E1; g.s. \to continuum)$ in
$^{11}$Li. Zero-width peaks with diamonds on the top (1)
given in arbitrary units, have been obtained
in the pure variational approach for both the ground and exited states, i.e.
the wave functions for the initial and the final states have been
obtained by the diagonalization of the truncated Hamiltonian matrix.
The curve with dots (2) has been calculated with the variational ground
state wave function
while the continuum spectrum wave function has been obtained
by the $J$-matrix approach. The solid curve (3) is the result of $J$-matrix
calculations for both the ground and continuum spectrum states.}
\end{figure}
\par
The reduced probability of the $E1$ transition obtained in the cluster
model, \linebreak
${\cal B}(E1;\;E_{f}-E_{0})$, is displayed on the figure 5.
In the $J$-matrix approach, zero-width pure variational peaks of ${\cal
B}(E1;\;E_{f}-E_{0})$ gain finite widths and shift to lower energies.  It
is seen that in the
calculation of ${\cal B}(E1;\;E_{f}-E_{0})$ it is important
to allow for the asymptotics not only for continuum states, but for the
ground state as well. Due to the wide spatial distribution of the halo
neutrons density,  large distances contribute essentially to the $E1$
strength in the low-energy region. To account for this contribution and to
get a convergence,  one should allow  for
the ground and continuum state wave
function components with very large values of the total number of
oscillator quanta $N \simeq 2000$\footnote{The classical turning point for
the basis function with the total number of oscillator quanta $N=1000$ is
at a distance of $\approx$108 fm from the origin.} in calculations of
${\cal B}(E1;\;E_{f}-E_{0})$ .  The large-distance $E1$ transitions enhance
the peak of the cluster ${\cal B}(E1;\;E_{f}-E_{0})$ and shift it to a
lower energy. Obviously, this peak should be associated with  the soft
dipole mode. Thus, the neutron halo manifests itself in the appearance of
the soft dipole mode that arises from the shift and enhancement of the
${\cal B}(E1;\;E_{f}-E_{0})$ peak.  This effect appears to be very
important in the calculation of the $^{11}$Li electromagnetic dissociation
cross section (see below).
\par
Large distances contribute essentially to $E2$ and $E0$ transitions, too.
The shift and the enhancement of  ${\cal B}(E0;\;E_{f}-E_{0})$ and of
${\cal B}(E2;\;E_{f}-E_{0})$ are even more pronounced. Nevertheless,
$E2$ and $E0$ transitions do not play an important role in the
electromagnetic dissociation, and we shall not discuss them here.
\begin{figure}[tbh]
\epsfverbosetrue
\epsfbox{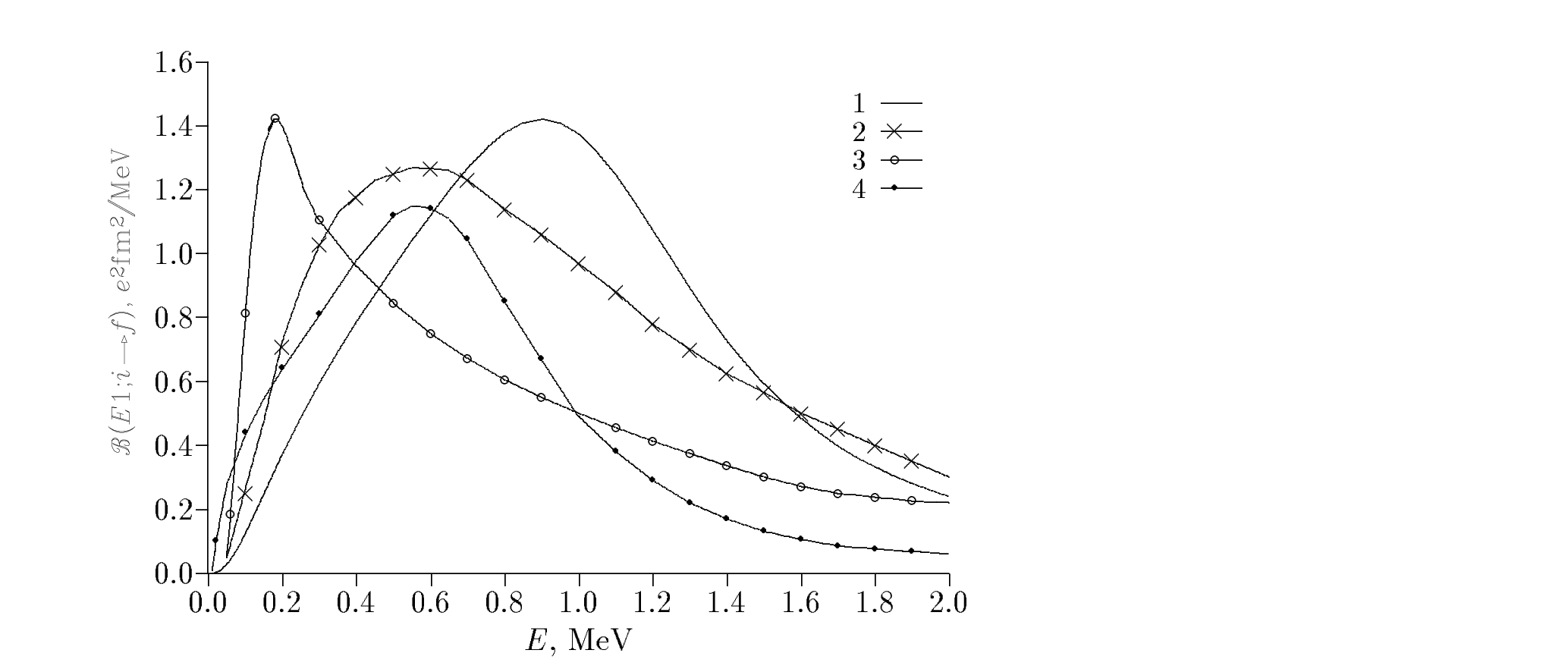}
\caption{ Comparison of our results$^{16}$ for ${\cal B}(E1; g.s. \to
continuum)$ in $^{11}$Li with results of other authors. 1 ---
$J$-matrix method$^{16}$; 2 --- ref.~$^{21}$; 3 --- ref.~$^{22}$;
4 --- experimental data parametrization of ref.~$^{20}$.}
\end{figure}
\par
Figure 6 shows the comparison of the results of our calculations$^{16}$ of
the cluster ${\cal B}(E1;\;E_{f}-E_{0})$
with the parametrization of experimental data of
ref.~$^{20}$. The agreement is reasonable. The form of the
 ${\cal B}(E1;\;E_{f}-E_{0})$ peak is well reproduced, the discrepancy in
the position of the  ${\cal B}(E1;\;E_{f}-E_{0})$ maximum is supposed
to be eliminated by the adjustment of  the
potentials. The results of the  ${\cal B}(E1;\;E_{f}-E_{0})$ calculations of
refs.~$^{21,22}$ are also depicted on fig.\ 6. The three-body
cluster calculations with the allowance for democratic decay channels of
ref.~$^{21}$ performed using usual coordinate-space hyperspherical
harmonics approach, nicely reproduce the energy of the soft dipole mode.
Nevertheless, the form of the peak in our calculations is reproduced
better. Note, that the authors of ref.~$^{21}$ used another
set of the potential parameters. The calculations of ref.~$^{22}$
with the allowance for two-body decay channels only, failed to
reproduce the form of the ${\cal B}(E1;\;E_{f}-E_{0})$ peak and
underestimate the energy of the soft dipole mode..
\par
The soft dipole mode exhausts about 90\%
of the cluster EWSR, $\sigma_{\infty}^{clust}$, associated with the
excitation of the cluster degrees of freedom only.  Nevertheless, it is
easy to show that
\begin{equation}
\frac{\sigma_{\infty}^{clust}}{\sigma_{\infty}^{tot}} =
\frac{1}{12} \approx 8.3\%\, ,     \label{14}
\end{equation}
where $\sigma_{\infty}^{tot}$ is the total EWSR
accounting for excitations of all nucleons.
So, the contribution from the soft dipole mode to the total
EWSR is relatively small. In the vicinity of the
${\cal B}(E1;\;E_{f}-E_{0})$ maximum
at the excitation energy $E\approx 1$ MeV only $\sim$8\% of the total EWSR
is exhausted. Nevertheless,
the account for the soft dipole mode  results in an essential
increase of the electromagnetic dissociation cross section
of 0.8 GeV/nucleon $^{11}$Li beams on Pb and Cu targets.
The wide space distribution of the halo neutrons density
is also
well-manifested in the electromagnetic dissociation
of $^{11}$Li beam.
The contribution of the large-distance $E\lambda$-transitions
to the cross section is about 50\%. The
calculations have been performed by the equivalent
photon method. The only parameter of the
method is the minimal value of the impact parameter $b_{min}$.
We use for $b_{min}$ the values of 9.0 fm for Pb and 6.8 fm for Cu target
nuclei, respectively.
These quantities are the sums of the $^{11}$Li and target nucleus charge
radii. With these values of $b_{min}$ we obtain for the
electromagnetic dissociation cross sections the
values of 0.966 barn for the Pb target and 0.132 barn for the Cu target; the
corresponding experimental values are $0.890\pm 0.110$ barn and
0.21$\pm$0.04 barn, respectively$^{23}$.
$E0$ and $E2$ transitions give only 1.2\%
contribution in the cross sections.
\vspace{3mm}
\par
\noindent
{\bf Acknowledgments}\\
\par
I am thankful to Yu.A.Lurie, T.Ya.Mikhelashvili and Yu.F.Smirnov for a
fruitful collaborate work some results of which have been
discussed above. Valuable discussions with J.M.Bang, B.V.Danilin,
R.I.Jibuti, I.J.Thompson and  J.S.Vaagen are acknowledged.
\par
This paper has been presented as a talk at the conference Perspectives of
Nuclear Physics in the Late Nineties (Hanoi, March 1994). The support
of my participation the conference by the International Science Foundation
is acknowledged.
\section*{References}
\begin{itemize}
\item[1.] S.P.Merkuriev and L.D.Faddeev, {\em Quantum scattering theory
for systems of few bodies} (Nauka Publishes, Moscow,  1985).
\item[2.] R.I.Jibuti, {\em Elem.\ Part.\ and Atom.\ Nucl.} {\bf 14}
(1983) 741.
\item[3.] R.I.Jibuti and N.B. Krupennikova, {\em  Hyperspherical
harmonics method  in  quantum   mechanics   of   few   bodies}
(Metsniereba,   Tbilisi, 1984).
\item[4.] O.V.Bochkarev et al, {\em Nucl.\ Phys.} {\bf A505} (1989) 215.
\item[5.] O.V.Bochkarev et al, {\em Yad.\ Phys.\ (Sov.\ J.\ Nucl.\
Phys.)} {\bf 55} (1992) 1729.
\item[6.] B.V.Danilin et al, {\em Yad.\ Phys.\ (Sov.\ J.\ Nucl.\
Phys.)} {\bf 46} (1987) 427.
\item[7.] W.Eyrich et al, {\em Phys.\ Rev.} {\bf C36} (1987) 416.
\item[8.] V.V.Burov, V.N.Dostovalov, M.Kaschiev and K.V.Shitikova, {\em
J.\ Phys.} {\bf G7} \linebreak[2] (1981) 131.
\item[9.] T.Ya.Mikhelashvili, A.M.Shirokov and Yu.F.Smirnov, {\em J.\
Phys.} {\bf G16} (1990) 1241.
\item[10.] H.A.Yamani and L.Fishman,  {\em J.\ Math.\ Phys.}  {\bf16}
(1975) 410; G.F.Filippov, {\em Yad.\ Phys.\ (Sov.\ J.\ Nucl.\ Phys.)} {\bf
33}(1981) 928; Yu.I.Nechaev and Yu.F.Smirnov, {\em Yad.\ Phys.\ (Sov.\
J.\ Nucl.\ Phys.)} {\bf 35} (1982) 1385.
\item[11.] Yu.F.Smirnov and A.M.Shirokov,  {\em Preprint} {\bf
ITP-88-47P} (Kiev, 1988); A.M.Shi\-ro\-kov, Yu.F.Smirnov and S.A.Zaytsev,
{\em to be published}.
\item[12.] S.Ali and A.Bodmer, {\em Nucl.\ Phys.} {\bf 80} (1966) 99.
\item[13.] D.Lebrun et al, {\em Phys.\ Lett.} {\bf 97B} (1980) 358.
\item[14.] C.A.Bertulani, F.L.Canto and M.H.Hussein, {\em Phys.\ Rep.} {\bf
226} (1993) 281.
\item[15.] Zhukov M.V. et al, {\em Phys.\ Rep.}  {\bf 231} (1993) 151.
\item[16.] Yu.A.Lurie, A.M.Shirokov and Yu.F.Smirnov,
{\em nucl-th}/{\bf 9407011}; {\em to be published.}
\item[17.] L.Johansen, A.S.Jensen and P.G.Hansen, {\em Phys.\ Lett.}
{\bf 244B} (1990) 357.
\item[18.] J.R\'{e}vai, M.Sotona and J.\u{Z}ofka, {\em J.\ Phys.} {\bf G11}
(1985) 745; J.Mare\u{s}, {\em Czech. J.\ Phys.} {\bf B37} (1987) 665.
\item[19.] T.Kobayashi et al, {\em Phys.\ Rev.\ Lett.} {\bf 60} (1988) 2599.
\item[20.] D.Sackett et al, {\em Preprint} {\bf MSUCL} (Michigan State
University, 1993).
\item[21.] B.V.Danilin, M.V.Zhukov, J.S.Vaagen, I.J.Thompson and J.M.Bang,
{\em Preprint} {\bf NORDITA -- 93/34 N} (Copenhagen, 1994); {\em private
communication.}
\item[22.] G.F.Bertsch and H.Esbensen, {\em Ann.\ of Phys.} {\bf 209}
(1991) 327; H.Esbensen and G.F.Bertsch,
{\em Nucl.\ Phys.} {\bf A542} (1992) 310.
\item[23.] T.Kobayashi et al, {\em Phys.\ Lett.} {\bf B232} (1989) 51.
\end{itemize}
\end{document}